\documentclass[epj,nopacs]{svjour}

\usepackage{amsmath}
\usepackage{graphicx}

\newcommand*{\hc}[1]{{#1}^{\dag}}
\newcommand*{\angled}[1]{\left\langle #1 \right\rangle}
\newcommand*{\secref}[1]{Section~\ref{sec:#1}}
\newcommand*{\figref}[1]{Figure~\ref{fig:#1}}
\newcommand*{\Figref}[1]{Figure~\ref{fig:#1}}
\newcommand*{\Figsref}[1]{Figures~\ref{fig:#1}}
\newcommand*{\ximax}{{\xi}_{\mathrm{max}}}
\newcommand*{\ket}[1]{{\left| #1 \right\rangle}}
\newcommand*{\abs}[1]{\left| #1 \right|}
\newcommand{\ev}[1]{{\left\langle #1 \right\rangle}}

\newcommand{\ii}{\mathrm{i}}
\newcommand{\dd}{\mathrm{d}}
\newcommand{\quantity}[3]{#1\times 10^{#2}\;#3}

\newcommand{\per}[1]{#1^{-1}}
\newcommand{\second}{\text{s}}
\newcommand{\comm}[2]{\left[\, #1 , #2 \,\right]}
\newcommand{\etal}{et al.}

\begin{document}

\title{Polariton dynamics of a disordered three-cavity system of four-level atoms}
\author{Abuenameh Aiyejina \and Roger Andrews}
%
%
\institute{The Department of Physics, The University of the West Indies, St. Augustine, Trinidad and Tobago}
\date{}
%
\abstract{
The effect of disorder in the intensity of the driving laser on the dynamics of a disordered three-cavity system of four-level atoms is investigated. This system can be described by a Bose-Hubbard Hamiltonian for dark-state polaritons. We examine the evolution of the first- and second-order correlation functions, the photon and atomic excitation numbers and the basis state occupation probabilities. We use the full Hamiltonian and the approximate Bose-Hubbard Hamiltonian with uniform and speckle disorder, as well as with different dipole couplings. We find that the results for the two Hamiltonians are in good agreement. We also find and that it is possible to obtain bunching and antibunching of the polaritons by varying the dipole couplings and that polaritons can be driven into a purely photonic state by varying the laser intensity.
}

\maketitle

\section{Introduction\label{sec:intro}}

The Bose-Hubbard model has received significant theoretical and experimental study as a model of quantum many-body phenomena since the seminal paper by Fisher \etal~\cite{Fisher:1989il}. This model has been experimentally realized using arrays of Josephson junctions~\cite{Fazio:2001fk} and cold atoms in optical lattices~\cite{Bloch:2008uq}. A great deal of work has focused on atoms in optical lattices since this system provides a defect-free lattice and control over the ratio of the on-site repulsion to the hopping strength via the laser intensity~\cite{Jaksch:1998xw}.

A variety of techniques have been used to study the dynamics of the Bose-Hubbard model. A review of some of these techniques in the context of out-of-equilibrium dynamics is given by Kennett~\cite{Kennett:2013ty}. These techniques include Bogoliubov, mean field and slave particle approaches, as well as numerical methods such as time-dependent density matrix renormalization group (t-DMRG), time evolving block decimation (TEBD), and exact diagonalization.

The addition of disorder to the Bose-Hubbard model leads to new phenomena such as the appearance of the Bose glass phase in addition to the Mott-insulator and superfluid phases of the clean model.
In this paper, we will examine the effect of disorder on the dynamics of a system proposed by Hartmann \etal~\cite{Hartmann:2006sj}. This system consists of an array of coupled optical cavities, each containing a large number of four-level atoms that are driven by an external laser with uniform intensity across the cavities. A brief review of the properties of coupled quantum electrodynamics cavities was done by Tomadin and Fazio~\cite{Tomadin:2010nr}. Hartmann \etal found that under certain conditions this system can be described by a Bose-Hubbard model for combined atom-photon excitations called dark-state polaritons and that it exhibits a Mott-insulator-to-superfluid transition that can be controlled using the intensity of the external laser. 

In the context of electromagnetically induced transparency (EIT), dark-state polaritons have been proposed as a method of stopping and storing light~\cite{Fleischhauer:2005ul,Lukin:2003ul,Fleischhauer:2000nr,Lukin:2000rz}. By tuning the intensity of the external laser, the photonic component of the polaritons can be mapped onto the atomic component, thereby stopping the propagation of photons. Such systems have been implemented using cold clouds of sodium atoms~\cite{Liu:2001qr} and hot rubidium vapor~\cite{Phillips:2001kx}. Another technique for storing light involves using ultracold rubidium atoms in a Mott-insulating state in a three-dimensional optical lattice~\cite{Schnorrberger:2009kl}.

The polariton system studied in this paper has the potential to be used as a quantum memory for the reversible storage and retrieval of photon states in an EIT scheme. This is similar to the scheme used with three-level $\Lambda$–-type atoms~\cite{Lukin:2003ul}. Quantum memories like this are required in quantum information processing protocols. In the polariton system, the Mott-insulator phase and the superfluid phase can be used as a basis for a storage scheme and a retrieval scheme respectively. The Mott-insulator phase not only localizes polaritons at each cavity, but also stores photons for a time that is determined by the lifetime of the polaritons. This lifetime is determined by the cavity decay rate and the decay rate of the atomic excitations~\cite{Hartmann:2006sj}. For a quantum memory using this system, the Rabi frequency of the driving laser can be used to reversibly switch the device between storing photons (Mott-insulator phase) and retrieving photons (superfluid phase).

The dynamics of three-cavity systems coupled to two-level systems have been examined by Felicetti \etal~\cite{Felicetti:2014fk} and Zhong \etal~\cite{Zhong:2012yq}. Felicetti \etal investigate photon transfer in three coupled cavities where the central one is coupled to a two-level system. Starting from an initial state with a photon in the leftmost cavity and the two-level system in the ground state, they find that inhomogeneity in the array coupling prevents complete transfer of the photon to the rightmost cavity in the strong-coupling regime. In contrast, they find that when the coupling between the cavity and the two-level system is in the ultrastrong-coupling regime, complete transfer is then possible. Zhong \etal consider a system of three cavities, each containing a two-level atom and show that by adjusting the atom-cavity detuning, cavity-cavity hopping rate and initial atomic states (for example, one atom in the excited state and two atoms in the ground state), a wide variety of time-evolution behaviours can be realized. In these two papers, open boundary conditions were used where there is no hopping of photons between the leftmost and rightmost cavities.
In addition, the effect of non-uniform laser intensities on the phase diagram and dynamics of the system of Hartmann \etal was investigated by Aiyejina and Andrews~\cite{Aiyejina:2016dp} using a Gutzwiller approach. For a one-dimensional system of 25 cavities, disorder in the laser intensity was found to lead to the appearance of a Bose-glass phase in the phase diagram. The evolution of the system from its ground state while subject to a ramp in the intensity of the laser was also considered. The addition of disorder in the laser intensity was found to affect the behaviour of the defect density, superfluid order parameter and excess energy pumped into the system.

In this paper, we consider a system composed of three cavities, each containing an ensemble of four-level atoms. We use periodic boundary conditions where there is photon hopping between the leftmost and rightmost cavities and we fix the cavity-cavity photon hopping rate. The initial state consists of one polariton in each cavity. We examine the effect of using a non-uniform intensity of the laser to drive the Mott-insulator-to-superfluid transition. We investigate the time evolution of the normalized first- and second-order correlation functions using the full Hamiltonian and the Bose-Hubbard Hamiltonian, both with and without disorder in the Rabi frequency. We also calculate the photon number and level 2 atomic excitation number in each cavity as a function of time for the full Hamiltonian, both with and without disorder. We also calculate the occupation probabilities of basis states for the full Hamiltonian that contain from 0 to 3 photons.

In \secref{cleanpolham} we introduce the full Hamiltonian for the polariton system and the approximate Bose-Hubbard Hamiltonian that describes dark state polaritons in this system. Then in \secref{disorderpolham} we show the extension of both models to a time-dependent, disordered Rabi frequency. In \secref{methods} we describe the methods used to perform the time evolution of the system and to determine the evolution of the  first- and second-order correlation functions and various basis states of the system. In \secref{results} we report the results obtained without disorder and the results obtained using both uniformly distributed and speckle disorder in the Rabi frequency.

\section{Theory\label{sec:theory}}

\subsection{The Polariton Hamiltonians of the Clean System\label{sec:cleanpolham}}

As described by Hartmann \etal~\cite{Hartmann:2006sj}, the polariton system consists of an array of cavities, each containing an ensemble of $N$ four-level atoms. Each cavity supports an electromagnetic mode with frequency $\omega$, and the cavities are sufficiently close together that the modes of adjacent cavities overlap. The overlap integral of the modes of adjacent cavities is given by $\alpha$. The full Hamiltonian of the polariton system is given by

\begin{align}
H_{\mathrm{Full}} = &\sum_{i}\bigg[\varepsilon\hc{S}_{12,i}S_{12,i} + \delta\hc{S}_{13,i}S_{13,i} + (\Delta + \varepsilon)\hc{S}_{14,i}S_{14,i}\notag\\
&+ \bigg(\Omega\hc{S}_{12,i}S_{13,i} + g_{13}\hc{a}_{i}S_{13,i}\notag\\
&+ g_{24}\hc{S}_{12,i}S_{14,i}\hc{a}_{i} + \mathrm{h.c.}\bigg)\bigg]\notag\\
&+ 2\omega\alpha\sum_{\angled{ij}}\left(\hc{a}_{i}a_{j} + \mathrm{h.c.}\right)\label{eq:Full}
\end{align}
where $\hc{a_i}$($a_i$) is the creation(annihilation) operator for photons in cavity $i$, and $\hc{S}_{12,i}$($S_{12,i}$), $\hc{S}_{13,i}$($S_{13,i}$) and\linebreak $\hc{S}_{14,i}$($S_{14,i}$) are the creation(annihilation) operators for collective atomic excitations from level 1 to levels 2, 3 and 4 respectively for cavity $i$. The dipole couplings between the cavity mode and the 1-3 and 2-4 transitions are given by $g_{13}$ and $g_{24}$ respectively. The Rabi frequency of the driving of the 2-3 transition by the external laser is given by $\Omega$. Finally, $\varepsilon$, $\delta$ and $\Delta$ are detunings from levels 2, 3 and 4 respectively.

Hartmann \etal~\cite{Hartmann:2006sj} show that under certain conditions the Hamiltonian \eqref{eq:Full} can be approximated by a Bose-Hubbard Hamiltonian for dark-state polaritons that is given by
\begin{equation}
H_{\mathrm{BH}} = U\sum_{i}{n_i\left(n_i - 1\right)} + J\sum_{\angled{ij}}{\left(\hc{p}_ip_j + \hc{p}_jp_i\right)},\label{eq:BH}
\end{equation}
where the on-site repulsion and hopping strength are given by
\begin{equation}
U = -\frac{g_{24}^2}{\Delta}\frac{g^2\Omega^2}{\left(g^2 + \Omega^2\right)^2}
\quad \text{and} \quad
J = \frac{2\omega\alpha\Omega^2}{g^2 + \Omega^2}
\end{equation}
respectively. The dark-state polariton creation operator is given by
\begin{equation}
	\hc{p}_{i} = \frac{1}{B}\left(g\hc{S}_{12,i} - \Omega\hc{a}_i\right),
\end{equation}
where we have used the substitutions $g = \sqrt{N}g_{13}$ and $B = \sqrt{g^2 + \Omega^2}$. The polariton number operator is given by $n_i = \hc{p}_i p_i$. In this model the on-site repulsion and hopping strength are coupled through the parameters $g_{13}$ and $\Omega$.

This Bose-Hubbard Hamiltonian can be obtained by first diagonalizing the on-site, atomic part of the full Hamiltonian in the absence of coupling to level 4, i.e with $g_{24} = 0$. This process results in a Hamiltonian containing creation and annihilation operators for three species of polaritons, one of which is the dark-state polariton that involves only operators for the photonic and level 2 atomic excitations. In addition to the photonic and level 2 atomic excitations, the other two polariton species also involve operators for the level 3 atomic excitations. The on-site repulsion term in the Bose-Hubbard Hamiltonian is obtained by adding the coupling to level 4 pertubatively to the diagonalized Hamiltonian. The hopping term is then obtained by rewriting the photon hopping term using the dark-state polariton operators and then ignoring high-energy terms involving the other polariton operators.

In this paper, we first examine the time evolution of the normalized first- and second-order correlation functions. The normalized first-order and second-order correlation functions are given by 
\begin{equation}
g_{i,j}^{(1)} = \frac{\ev{\hc{p_i} p_j}}{\sqrt{\ev{n_i}\ev{n_j}}} \quad \textrm{and} \quad g_{i,j}^{(2)} = \frac{\ev{\hc{p_i}\hc{p_j}p_j p_i}}{\ev{n_i}\ev{n_j}}
\end{equation}
respectively.
In the simulations we neglect losses from the cavity mode and atomic level 2, since these are negligible over the time period considered in the calculations~\cite{Hartmann:2006sj}.

\subsection{The Polariton Hamiltonians of the Disordered System\label{sec:disorderpolham}}

In the case of a time-dependent Rabi frequency with disorder, the full Hamiltonian of the system becomes
\begin{align}
H_{\mathrm{Full}}(t) = &\sum_{i}\bigg[\varepsilon\hc{S}_{12,i}S_{12,i} + \delta\hc{S}_{13,i}S_{13,i} + (\Delta + \varepsilon)\hc{S}_{14,i}S_{14,i}\notag\\
&+ \bigg(\Omega_i(t)\hc{S}_{12,i}S_{13,i} + g_{13}\hc{a}_{i}S_{13,i}\notag\\
&+ g_{24}\hc{S}_{12,i}S_{14,i}\hc{a}_{i} + \mathrm{h.c.}\bigg)\bigg]\notag\\
&+ 2\omega\alpha\sum_{\angled{ij}}\left(\hc{a}_{i}a_{j} + \mathrm{h.c.}\right),\label{eq:Fulldis}
\end{align}
where $\Omega_i(t)$ is the time-dependent Rabi frequency at cavity $i$.
Meanwhile, the Bose-Hubbard Hamiltonian of the system becomes
\begin{equation}
H_{\mathrm{BH}}(t) = \sum_{i}{U_i(t) n_i\left(n_i - 1\right)} + \sum_{\angled{ij}}{J_{ij}(t)\left(\hc{p}_{i}p_{j} + \hc{p}_{j}p_{i}\right)},\label{eq:BHdis}
\end{equation}
where $U_i(t)$ is the time-dependent strength of the on-site repulsion at cavity $i$ and $J_{ij}(t)$ is the time-dependent strength of the hopping between cavities $i$ and $j$. The on-site repulsion and hopping strength are now given by
\begin{subequations}
\begin{equation}
U_i(t) = -\frac{g_{24}^2}{\Delta}\frac{g^2\Omega_i(t)^2}{\left(g^2 + \Omega_i(t)^2\right)^2}
\end{equation}
and
\begin{equation}
J_{ij}(t) = \frac{2\omega\alpha\Omega_i(t)\Omega_j(t)}{\sqrt{g^2 + \Omega_i(t)^2\vphantom{\Omega_j(t)^2}}\sqrt{g^2 + \Omega_j(t)^2}}.
\end{equation}
\end{subequations}
The cavity- and time-dependent dark-state polariton creation operator is given by
\begin{equation}
\hc{p}_i(t) = \frac{1}{\sqrt{g^2 + \Omega_i(t)^2}}\left(g\hc{S}_{12,i} - \Omega_i(t)\hc{a}_i\right).
\end{equation}

In this paper we consider both uniformly distributed disorder and speckle disorder in the Rabi frequency. For the uniform disorder, the Rabi frequency at cavity $i$ is given by $\Omega_i(t) = \Omega(t)(1 + \xi_i)$, where $\Omega(t)$ is the time-dependent mean Rabi frequency and $\xi_i$ is an uncorrelated random variable uniformly distributed in the interval $\left[-\ximax, \ximax\right]$. While the uncorrelated, uniformly distributed disorder is used as a starting point, correlated speckle disorder in the laser intensity is more realistic. Such disorder could be created, for example, by reflection of the laser off of a rough surface. The reflection of the laser off such a surface leads to an electric field with a complex amplitude that is the sum of many contributions scattered randomly from the surface. Under certain assumptions about the scattering process, the resultant complex amplitude of the electric field at a point away from the surface has real and imaginary parts that are uncorrelated and normally distributed. The amplitude of the resultant field therefore has a Rayleigh distribution. Since the Rabi frequency is proportional to the electric field amplitude, the Rabi frequency will also have a Rayleigh distribution. This process results in a disordered intensity pattern containing randomly distributed bright spots called "speckles". The values for the Rabi frequency with speckle disorder are obtained from intensity values calculated using the method of Duncan and Kirkpatrick~\cite{Duncan:2008kx}.

\section{Methods\label{sec:methods}}

For this paper, we performed the simulations of the time dynamics by building the time-dependent Hamiltonian matrix and performing numerical integration of the Schr\"{o}dinger equation. The matrix was constructed in a subspace of the Hilbert space with a particular fixed number of excitations.

In the case of the full Hamiltonian, we used a basis consisting of Fock states composed of the occupation numbers for the cavity mode and the collective atomic excitations. This basis is given by $\left\{\ket{\vec{n}_{1},\vec{n}_{2},\vec{n}_{3}} \mid \abs{\vec{n}_{1}} + \abs{\vec{n}_{2}} + \abs{\vec{n}_{3}} = 3\right\}$, where the occupation number vector for cavity $i$ is given by $\vec{n}_{i} = \left(n^{(\mathrm{ph})}_{i},n^{(2)}_{i},n^{(3)}_{i},n^{(4)}_{i}\right)$ and $\abs{\vec{n}_{i}} = n^{(\mathrm{ph})}_{i} + n^{(2)}_{i} + n^{(3)}_{i} + 2n^{(4)}_{i}$ is the number of excitations in cavity $i$. Here $n^{(\mathrm{ph})}_{i}$ is the photon occupation number for cavity $i$ and $n^{(2)}_{i}$, $n^{(3)}_{i}$ and $n^{(4)}_{i}$ are the atomic levels 2, 3 and 4 occupation numbers for cavity $i$ respectively. An atomic level 4 occupation was counted as two excitations since atomic level 4 has twice the energy of levels 2 and 3.

The matrices for the number-conserving operators\linebreak $\hc{S}_{12,i}S_{12,i}$, $\hc{S}_{13,i}S_{13,i}$, $\hc{S}_{14,i}S_{14,i}$, $\hc{S}_{12,i}S_{13,i}$, $\hc{a}_i S_{13,i}$,\linebreak $\hc{S}_{12,i}S_{14,i}\hc{a}_{i}$, $\hc{a}_i a_j$ and their Hermitian conjugates were constructed in this basis, and the matrix of the Hamiltonian $H_{\mathrm{Full}}$ was calculated using $\eqref{eq:Fulldis}$. Additionally, the operators $\hc{S}_{12,i}S_{12,j}$, $\hc{a}_i S_{12,j}$ and $\hc{S}_{12,i}a_j$ were calculated for use in expanding the operators $\hc{p}_ip_j$. The initial state for the simulations, $\ket{\psi_0}$, was obtained by expanding the product of the dark-state polariton creation operators for each cavity, $\hc{p}_1(0)\hc{p}_2(0)\hc{p}_3(0)$, in terms of the operators $\hc{a}_i$ and $\hc{S}_{12,i}$, and then replacing the products of these operators with the basis states that would be obtained by applying them to the vacuum state. This effectively gives $\ket{\psi_0} = \hc{p}_1(0)\hc{p}_2(0)\hc{p}_3(0)\ket{0,0,0}$. This process was necessary since the vacuum state does not belong to the subspace considered.

For the Bose-Hubbard Hamiltonian, we used the occupation number basis $\left\{\ket{n_{1},n_{2},n_{3}} \mid n_{1} + n_{2} + n_{3} = 3\right\}$, where $n_{i}$ is the polariton number for cavity $i$. The matrices for the number-conserving operators $n_i$ and $\hc{p}_{i}p_{j}$ were constructed in this basis, and the matrix of the Hamiltonian $H_{\mathrm{BH}}$ was calculated using $\eqref{eq:BHdis}$.

Once the Hamiltonian matrix was calculated, the matrix equation $\ii\frac{\dd}{\dd t}\ket{\psi(t)} = H\ket{\psi(t)}$ was expanded as a\linebreak system of ordinary differential equations for the time-\linebreak dependent elements of the vector $\ket{\psi(t)}$. This system was then solved using a Runge-Kutta method.
 
\section{Results\label{sec:results}}

\begin{figure*}
	\begin{center}
		\includegraphics[width=0.95\textwidth]{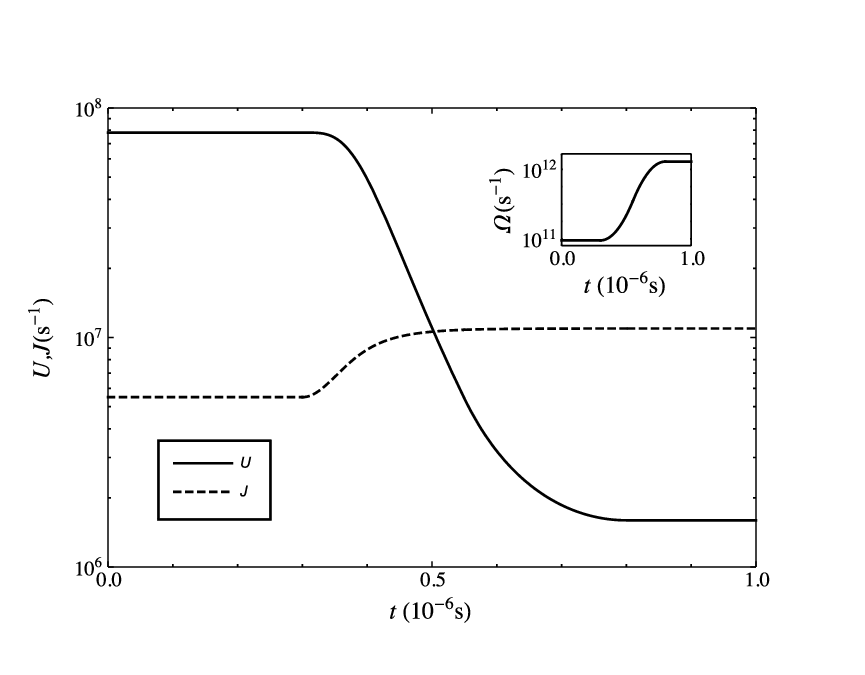}
	\end{center}
	\caption{The time-dependence of $\Omega$, $U$ and $J$.\label{fig:WJUplot}}
\end{figure*}

We consider a system of 3 cavities with periodic boundary conditions with an initial state of one polariton in each cavity. The Rabi frequency initially has a value of $\quantity{7.9}{10}{\per\second}$ and is increased to $\quantity{1.1}{12}{\per\second}$ over a time period of $0.5\,\mu\second$ as shown in the inset of \Figref{WJUplot}. The corresponding variations of $U$ and $J$ are also shown in \Figref{WJUplot}. This evolution is adiabatic since $gB^{-2}(\dd\Omega/\dd t) \ll \abs{\mu_{+}}, \abs{\mu_{-}}$, where $\mu_{\pm} = (\delta\pm \sqrt{4B^{2}+\delta^{2}})/2$ are the energies of the two species of polaritons that do not enter into the Bose-Hubbard Hamiltonian~\cite{Hartmann:2006sj}.

\subsection{No Disorder\label{sec:clean}}

For the simulations in this section, the system parameters used were $g_{13} = g_{24} = \quantity{2.5}{9}{\per\second}$, $\varepsilon = 0$, $\delta = \quantity{1.0}{12}{\per\second}$, $\Delta = \quantity{-2.0}{10}{\per\second}$, $N = 1000$, and $2\omega\alpha = -\quantity{1.1}{7}{\per\second}$. We study the case of no disorder and examine the time evolution of various correlation functions and of the occupation probabilities of certain states.

\subsubsection{Correlation Functions\label{sec:cleancorr}}

For both the full Hamiltonian and the Bose-Hubbard Hamiltonian we calculate the time evolution of the normalized first-order correlation between cavities 1 and 2, $g_{1,2}^{(1)}$, the normalized second-order correlation for cavity 1, $g_{1,1}^{(2)}$, and the normalized second-order correlation between cavities 1 and 2, $g_{1,2}^{(2)}$. Since the system is uniform, the values of $g_{i,j}^{(1)}$, $g_{i,i}^{(2)}$ and $g_{i,j}^{(2)}$ are the same for all other cavities and pairs of adjacent cavities.

\begin{figure*}
	\begin{center}
		\includegraphics[width=0.95\textwidth]{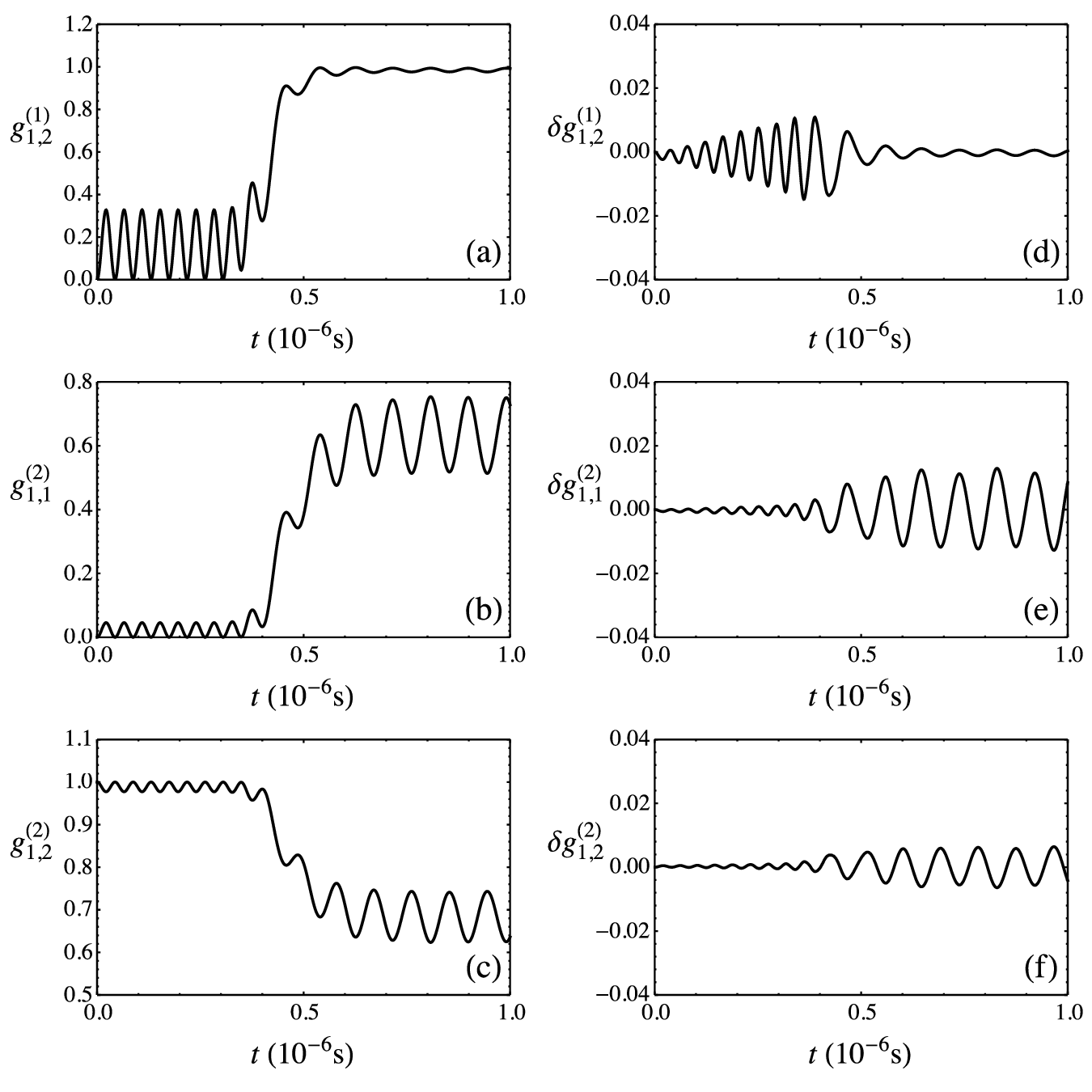}
	\end{center}
	\caption{The (a) normalized first-order correlation between cavities 1 and 2, (b) normalized second-order correlation in cavity 1 and (c) normalized second-order correlation between cavities 1 and 2. The results in (a), (b) and (c) were simulated using the full Hamiltonian in the absence of disorder. Also shown is the difference between the results for the full and Bose-Hubbard Hamiltonians for the (d) normalized first-order correlations, (e) normalized on-site second-order correlations and (f) normalized inter-cavity second-order correlations. The results in (d), (e) and (f) were obtained in the absence of disorder.\label{fig:fulldiffplot0}}
\end{figure*}

\Figsref{fulldiffplot0} (a), (b) and (c) show the results for the full Hamiltonian and \Figsref{fulldiffplot0} (d), (e) and (f) show the differences between the results for the full Hamiltonian and the approximate Bose-Hubbard Hamiltonian. For the latter half of the time evolution, the maximum percentage difference for the first-order correlation function is about 0.42\%, the maximum for the on-site second-order correlation function is about 2.1\% and the maximum for the inter-cavity second-order correlation function is about 0.95\%. The percentage differences for the first-order correlation function and on-site second-order correlation function are greater in the first half of the time evolution due to the smaller absolute values of the functions there that lead to increased sensitivity to the differences between the results. Nevertheless, the fact that the late time behaviour of both results agree well indicates that the Bose-Hubbard Hamiltonian gives a fairly good approximation to the full Hamiltonian, with respect to the correlation functions.

The results show that as the Rabi frequency is increased, the normalized first-order correlation function and the normalized on-site second-order correlation function increase and the normalized inter-cavity second-order correlation function decreases. The number fluctuations of polaritons in cavity $i$ is given by $F_i = \ev{n_i^2} - \ev{n_i}^2$. Since the total number of polaritons is 3 and there is no disorder, $\ev{n_i} = 1$ for all cavities. Therefore, $F_i = \ev{n_i^2} - 1$.
Due to the relationship $\comm{p_i}{\hc{p}_j} = \delta_{ij}$, we have $\ev{\hc{p_i}\hc{p_j}p_j p_i} = \ev{n_i n_j} - \delta_{ij}\ev{n_i}$ and
\begin{equation}
	g_{i,i}^{(2)} = \frac{\ev{\hc{p_i}\hc{p_i}p_i p_i}}{\ev{n_i}\ev{n_i}} = \frac{\ev{n_i^2} - \ev{n_i}}{\ev{n_i}^2} = \ev{n_i^2} - 1
\end{equation}
As a result, in the absence of disorder, the number fluctuations, $F_i$, are equal to the normalized on-site second-order correlation function, $g_{i,i}^{(2)}$. From \Figref{fulldiffplot0} (b), we see that the on-site second-order correlations/number fluctuations are initially small since the initial Rabi frequency puts the system in the Mott-insulator phase. There are initially some small oscillations because the initial state of one polariton in each cavity is not exactly the ground state of the initial Hamiltonian. As the Rabi frequency increases, the system is driven into the superfluid phase, which has large number fluctuations since the polaritons are able to move more freely between cavities. From \figref{fulldiffplot0} (a), we see that the normalized first-order correlation function, $g_{1,2}^{(1)}$, starts off oscillating between 0 and 0.33 in the Mott-insulator phase and then increases as the Rabi frequency increases to a value near 1 in the superfluid phase. From \figref{fulldiffplot0} (b), we see that the normalized on-site second-order correlation function, $g_{1,1}^{(2)}$, is always less than 1, indicating that there is antibunching of the polaritons. The normalized inter-cavity second-order correlation function, $g_{1,2}^{(2)}$, starts off at 1 and then decreases as the Rabi frequency increases.

\subsubsection{Photon and Collective Level 2 Excitation Numbers\label{sec:cleannum}}

\begin{figure*}
	\begin{center}
		\includegraphics[width=0.95\textwidth]{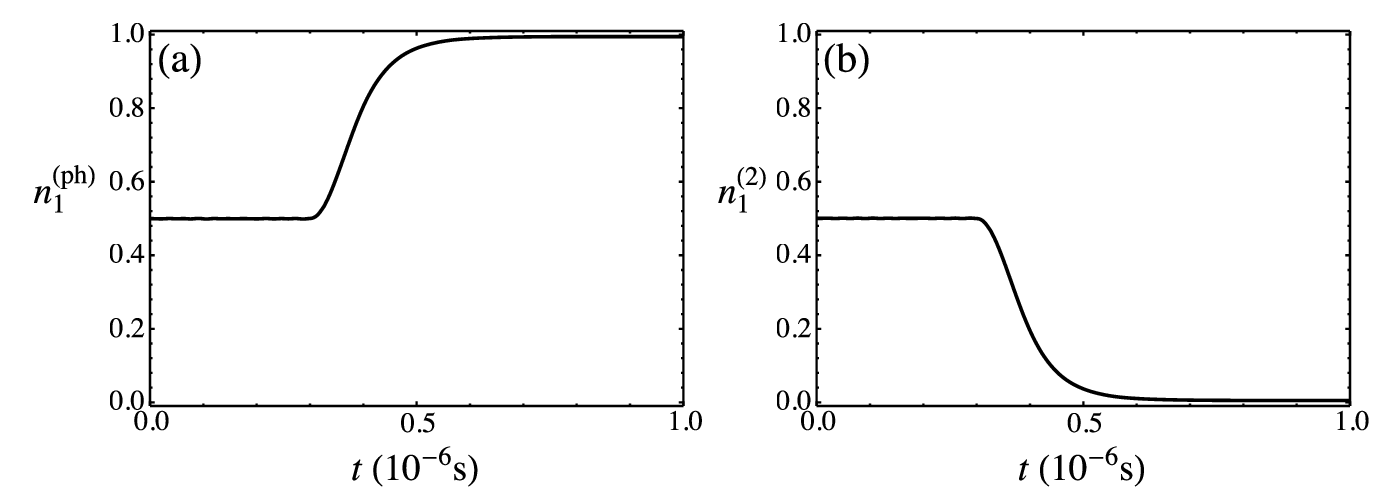}
	\end{center}
	\caption{The evolution of (a) the photon number in cavity 1 and (b) the collective level 2 excitation number in cavity 1.\label{fig:naS12plot0}}
\end{figure*}

\Figsref{naS12plot0} (a) and (b) show the evolution of the photon number for cavity 1, $n^{(\mathrm{ph})}_{1}$, and collective level 2 excitation number for cavity 1, $n^{(2)}_{1}$, respectively. In the absence of disorder, these values are the same for all cavities. At the beginning of the evolution, the photon number and atomic excitation number are roughly equal, having values of approximately 0.5 each. This is due to the fact that initially $g = \sqrt{N}g_{13} \approx \Omega = \quantity{7.9}{10}{\per\second}$, and the polariton operator is roughly symmetric in the photon and atomic excitation operators. If we were to have $g \gg \Omega$, then the atomic contribution to the polariton operator would dominate, while if $g \ll \Omega$, the photonic contribution would dominate. As the Rabi frequency increases, the photon number increases to a value of almost 1, while the atomic excitation number decreases to almost 0. We see that the polaritons become almost completely photonic by the end of the evolution, as would be expected in the superfluid phase since it is the photons that are responsible for the hopping between cavities.

\subsubsection{Occupation Probabilities\label{sec:cleanprob}}

\begin{figure*}
	\begin{center}
		\includegraphics[width=0.95\textwidth]{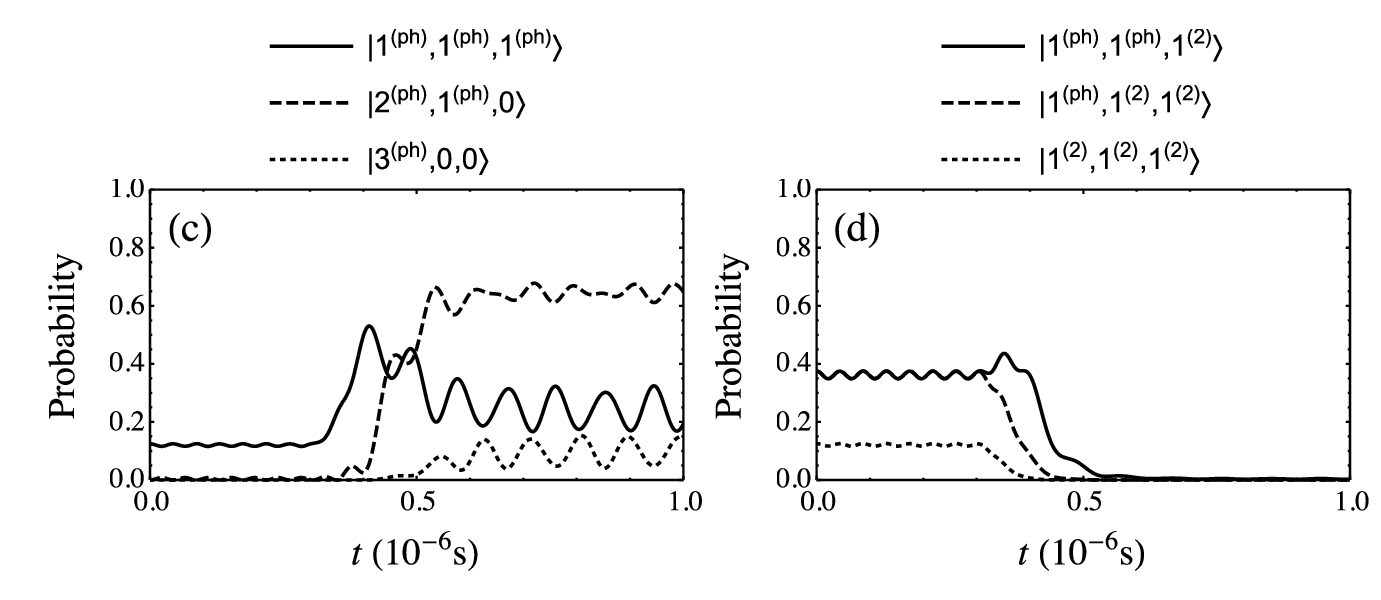}
	\end{center}
	\caption{The evolution of the occupation probability of 6 classes of basis states in the wavefunction.\label{fig:popsplot0}}
\end{figure*}

We further examine the occupation probabilities of certain basis states during the time evolution. \Figref{popsplot0} shows the results for 6 classes of states. For the purely photonic states, $\ket{1^{(\mathrm{ph})},1^{(\mathrm{ph})},1^{(\mathrm{ph})}}$ represents the state with 1 photon in each cavity, $\ket{2^{(\mathrm{ph})},1^{(\mathrm{ph})},0}$ represents the sum over states with 2 photons in one cavity and 1 photon in another and $\ket{3^{(\mathrm{ph})},0,0}$ represents the sum over states with 3 photons in one cavity. For the mixed states (those containing both photons and atomic excitations), $\ket{1^{(\mathrm{ph})},1^{(\mathrm{ph})},1^{(2)}}$ represents the sum over states with 1 photon each in two cavities and 1 collective excitation to atomic level 2 in the remaining cavity, $\ket{1^{(\mathrm{ph})},1^{(2)},1^{(2)}}$ represents the sum over states with 1 photon in one cavity and 1 collective excitation to level 2 in each of the remaining two cavities and $\ket{1^{(2)},1^{(2)},1^{(2)}}$ represents the state with 1 collective excitation to level 2 in each cavity. The occupation probabilities of all other states were found to be negligible during the system's evolution.

Since the initial state has one polariton in each cavity, each of the states having 1 photon/atomic excitation in each cavity starts off with a non-zero occupation probability, while the other states start off with zero probability. As the Rabi frequency is increased, the occupation probabilities of the mixed states decrease to 0, leading to the loss of the atomic part of the polaritons as previously noted. The occupation probabilities of the states having a cavity with 2 photons and the states having a cavity with 3 photons increase as the Rabi frequency is increased, while the probability of the state with 1 photon in each cavity initially increases to a maximum and then decreases. All the probabilities exhibit oscillations superimposed on the general increase or decrease. The initial increase in the probability of the $\ket{1^{(\mathrm{ph})},1^{(\mathrm{ph})},1^{(\mathrm{ph})}}$ state, as well as the decrease in the probabilities of the mixed states, occurs as the polaritons become photonic and the subsequent decrease in the probability of the $\ket{1^{(\mathrm{ph})},1^{(\mathrm{ph})},1^{(\mathrm{ph})}}$ state occurs as the probability is redistributed to the other photonic states.

\subsection{Uniform Disorder\label{sec:uniform}}

In order to investigate the effect of disorder, we first obtained results when uniform disorder was introduced into the Rabi frequency with disorder strengths of $\ximax$ = 0.1, 0.25 and 0.5. We used 400 realizations of uniform disorder. The values used for $g_{13}$, $g_{24}$, $\varepsilon$, $\delta$, $\Delta$, $N$ and $2\omega\alpha$ were the same as those used in section \ref{sec:clean}.

\subsubsection{Correlation Functions\label{sec:uniformcorr}}

\begin{figure*}
	\begin{center}
		\includegraphics[width=0.95\textwidth]{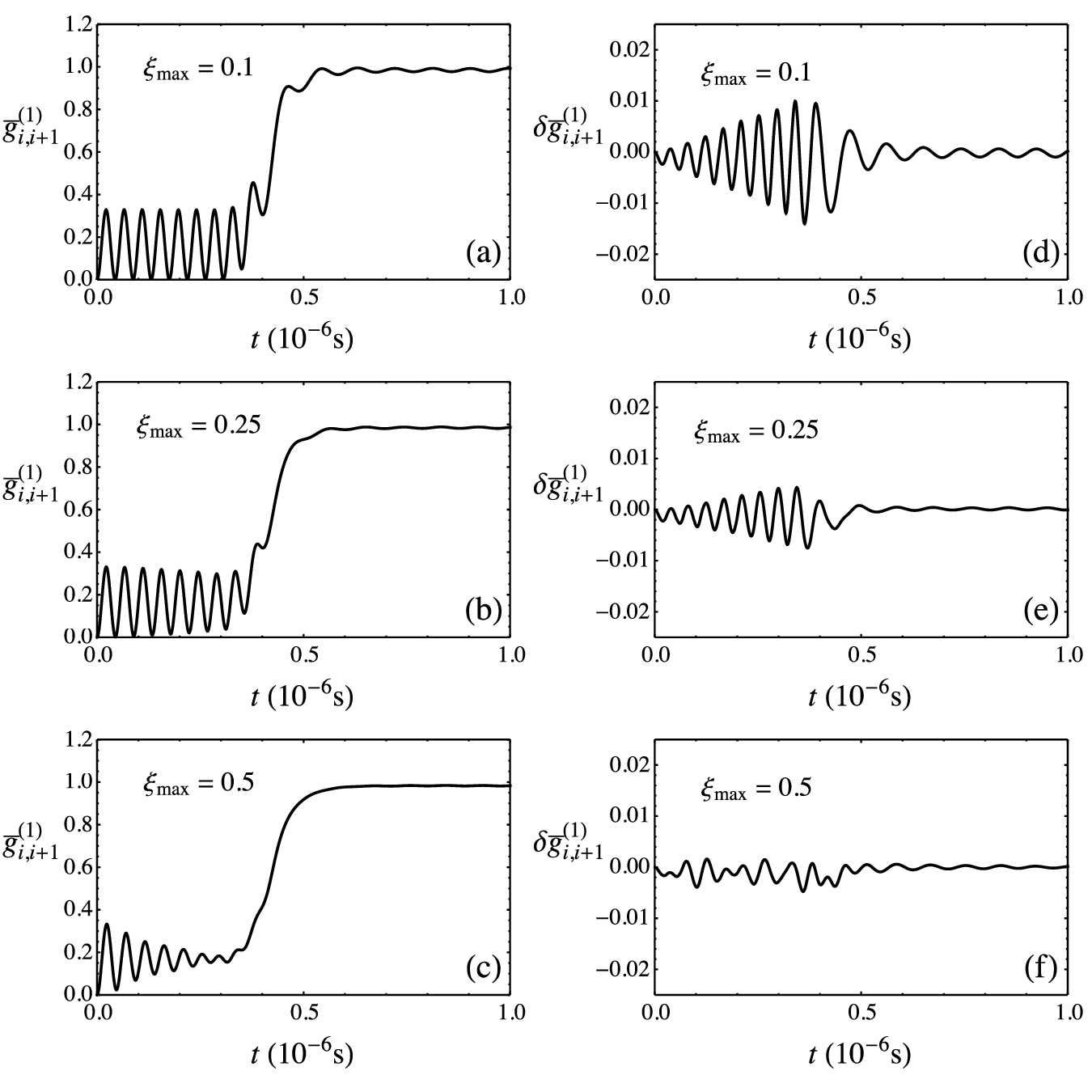}
	\end{center}
	\caption{The normalized first-order correlation between adjacent cavities averaged over all 3 cavities and 400 realizations of disorder for the full Hamiltonian with disorder strengths, $\ximax$, of (a) 0.1, (b) 0.25 and (c) 0.5. Also shown is the difference in the average normalized first-order correlation between the full and Bose-Hubbard Hamiltonians for $\ximax$ = (d) 0.1, (e) 0.25 and (f) 0.5.\label{fig:g1fulldiffplot}}
\end{figure*}

\begin{figure*}
	\begin{center}
		\includegraphics[width=0.95\textwidth]{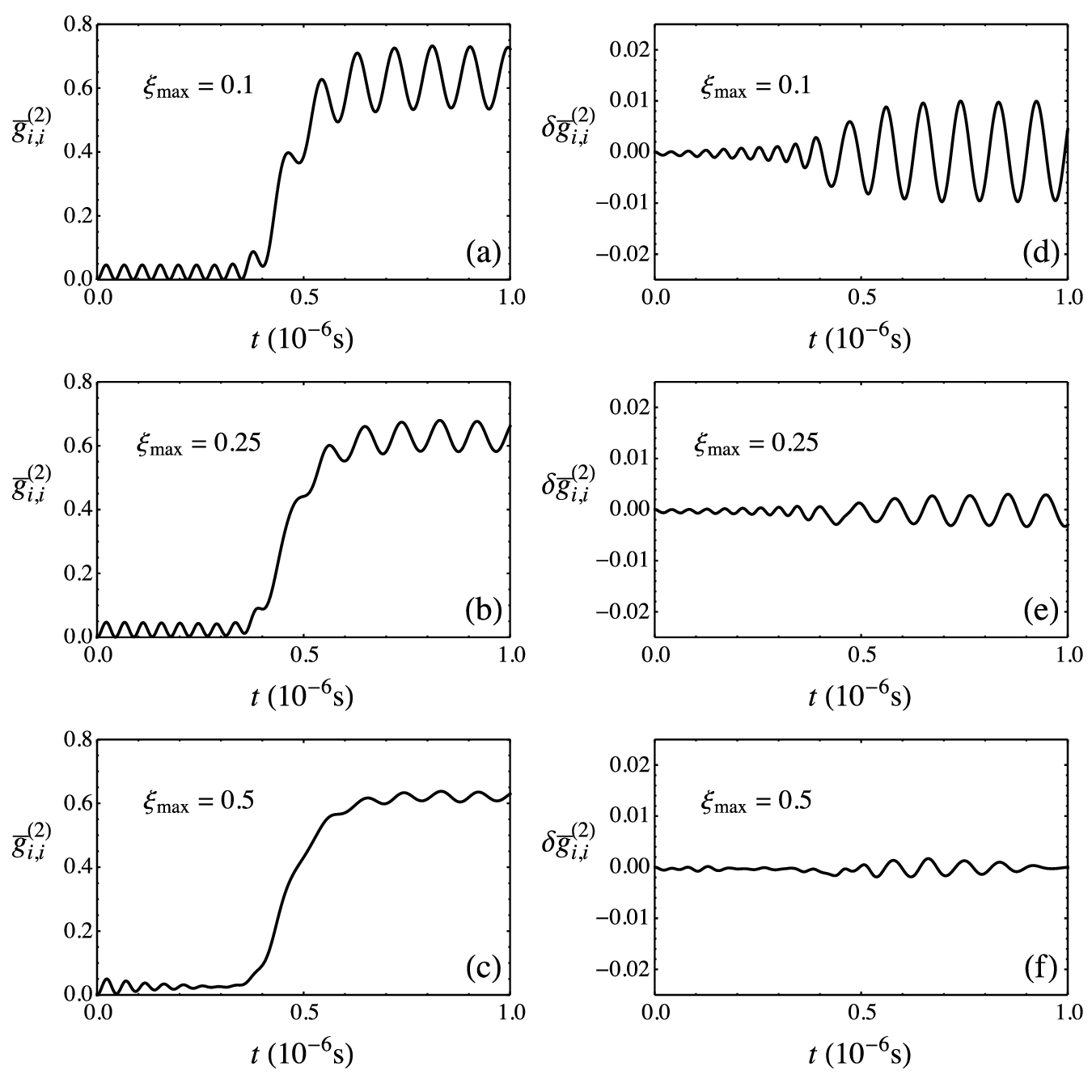}
	\end{center}
	\caption{The normalized on-site second-order correlation averaged over all 3 cavities and 400 realizations of disorder for the full Hamiltonian with disorder strengths, $\ximax$, of (a) 0.1, (b) 0.25 and (c) 0.5. Also shown is the difference in the average normalized on-site second-order correlation between the full and Bose-Hubbard Hamiltonians for $\ximax$ = (d) 0.1, (e) 0.25 and (f) 0.5.\label{fig:g2ifulldiffplot}}
\end{figure*}

\begin{figure*}
	\begin{center}
		\includegraphics[width=0.95\textwidth]{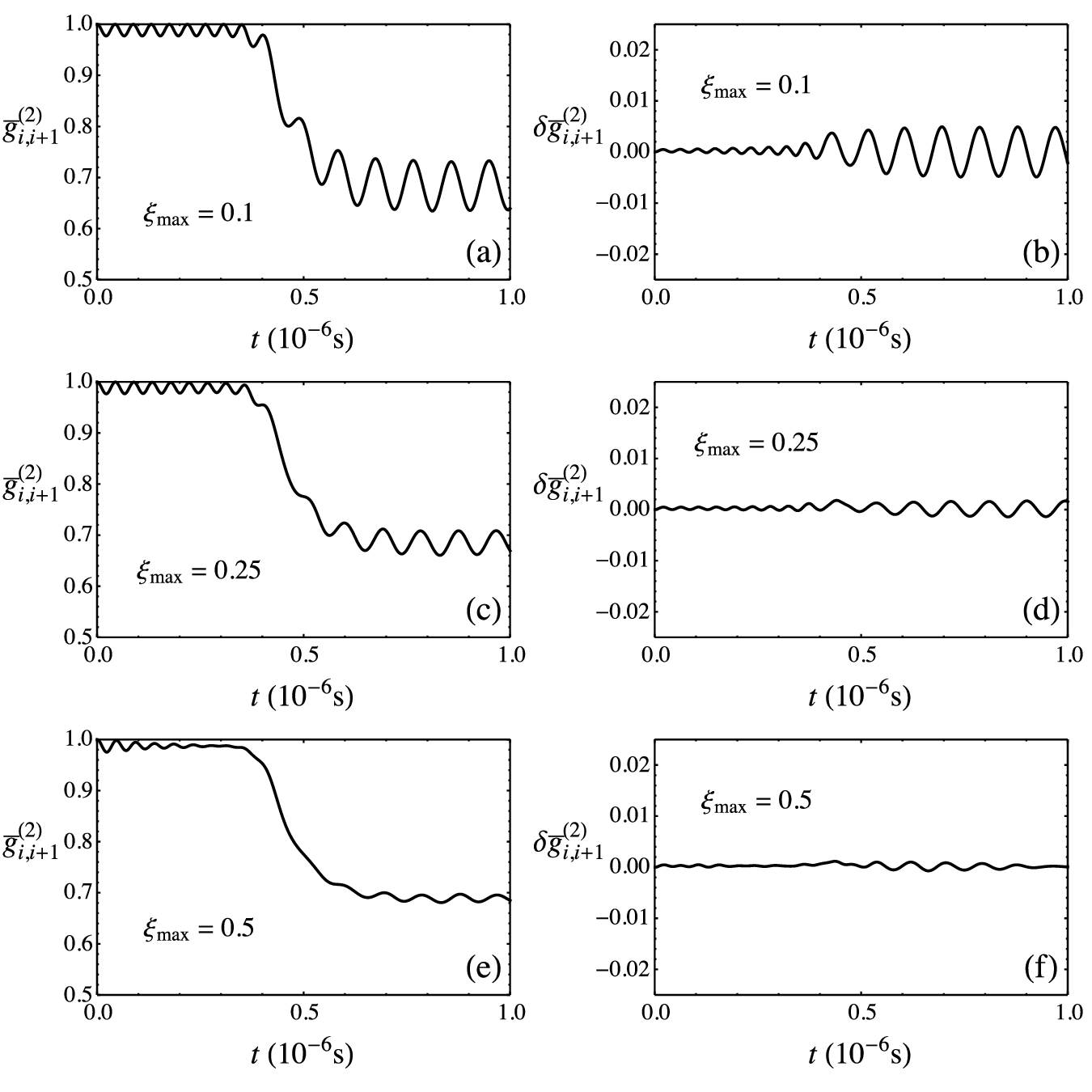}
	\end{center}
	\caption{The normalized second-order correlation between adjacent cavities averaged over all 3 cavities and 400 realizations of disorder for the full Hamiltonian with disorder strengths, $\ximax$, of (a) 0.1, (b) 0.25 and (c) 0.5. Also shown is the difference in the average normalized inter-cavity second-order correlation between the full and Bose-Hubbard Hamiltonians for $\ximax$ = (d) 0.1, (e) 0.25 and (f) 0.5.\label{fig:g2ijfulldiffplot}}
\end{figure*}

The number of realizations used was found to be sufficient for convergence of the results since the percentage change in the first-order correlations, on-site second-order correlations and inter-cavity second-order correlations upon going from 399 to 400 realizations in the average was less than 0.3\%, 0.3\% and 0.03\% respectively for both the full and Bose-Hubbard Hamiltonians. The average on-site second-order correlation functions were found by averaging over all 3 cavities and the 400 disorder realizations, while the average first-order and average inter-cavity second-order correlation functions were found by averaging over all pairs of adjacent cavities and the 400 disorder realizations.

\Figref{g1fulldiffplot} shows the results for the average first-order correlations. \Figsref{g1fulldiffplot} (a), (b) and (c) show the average first-order correlations for the three disorder strengths using the full Hamiltonian, while \Figsref{g1fulldiffplot} (d), (e) and (f) show the corresponding differences in the average first-order correlations between the results for the full Hamiltonian and the Bose-Hubbard Hamiltonian. The overall trend in the correlation function is the same as in the absence of disorder, but as the strength of the disorder increases, the oscillations in the correlation function are increasingly damped. For the largest disorder strength considered, the oscillations in the latter half of the time evolution are almost completely suppressed. For the latter half of the time evolution, the maximum percentage differences between the results for the full Hamiltonian and the Bose-Hubbard Hamiltonian are about 0.37\%, 0.08\% and 0.1\% for the disorder strengths 0.1, 0.25 and 0.5 respectively. Since the percentage difference in the absence of disorder was 0.42\%, the percentage difference decreases as the strength of the disorder increases, with the exception of the largest disorder strength.

Next, \Figref{g2ifulldiffplot} shows the results for the average on-site second-order correlations. \Figsref{g2ifulldiffplot} (a), (b) and (c) show the results using the full Hamiltonian for the three disorder strengths, while \Figsref{g2ifulldiffplot} (d), (e) and (f) show the differences between the results for the full and Bose-Hubbard Hamiltonians. As was the case with the first-order correlation function, the overall trend in the on-site second-order correlation function is the same as it was in the absence of disorder. Again, as the strength of the disorder increases, the oscillations in the correlation function are increasingly damped. For the latter half of the time evolution, the maximum percentage differences between the results for the two Hamiltonians are about 1.7\%, 0.5\% and 0.4\% for the disorder strengths 0.1, 0.25 and 0.5 respectively. Since the percentage difference in the absence of disorder was 2.1\%, the percentage difference decreases as the strength of the disorder increases.

Finally, \Figref{g2ijfulldiffplot} shows the results for the average inter-cavity second-order correlations. \Figsref{g2ijfulldiffplot} (a), (b) and (c) show the results using the full Hamiltonian for the three disorder strengths, while \Figsref{g2ijfulldiffplot} (d), (e) and (f) show the differences between the results for the full and Bose-Hubbard Hamiltonians. As before, the overall trend is the same as in the absence of disorder and the amplitude of the oscillations in the correlation function decreases as the strength of the disorder increases. For the latter half of the time evolution, the maximum percentage differences between the results for the two Hamiltonians are about 0.72\%, 0.25\% and 0.14\% for the disorder strengths 0.1, 0.25 and 0.5 respectively. Since the percentage difference in the absence of disorder was 0.95\%, the percentage difference again decreases as the strength of the disorder increases.

The results above show that the percentage differences between the results for the two Hamiltonians for the first- and second-order correlation functions are smaller for the three disorder strength as compared to the results in the absence of disorder. Therefore the Bose-Hubbard Hamiltonian provides a good approximation even in the presence of disorder. 

\subsubsection{Occupation Probabilities\label{sec:uniformprob}}

\begin{figure*}
	\begin{center}
		\includegraphics[width=0.95\textwidth]{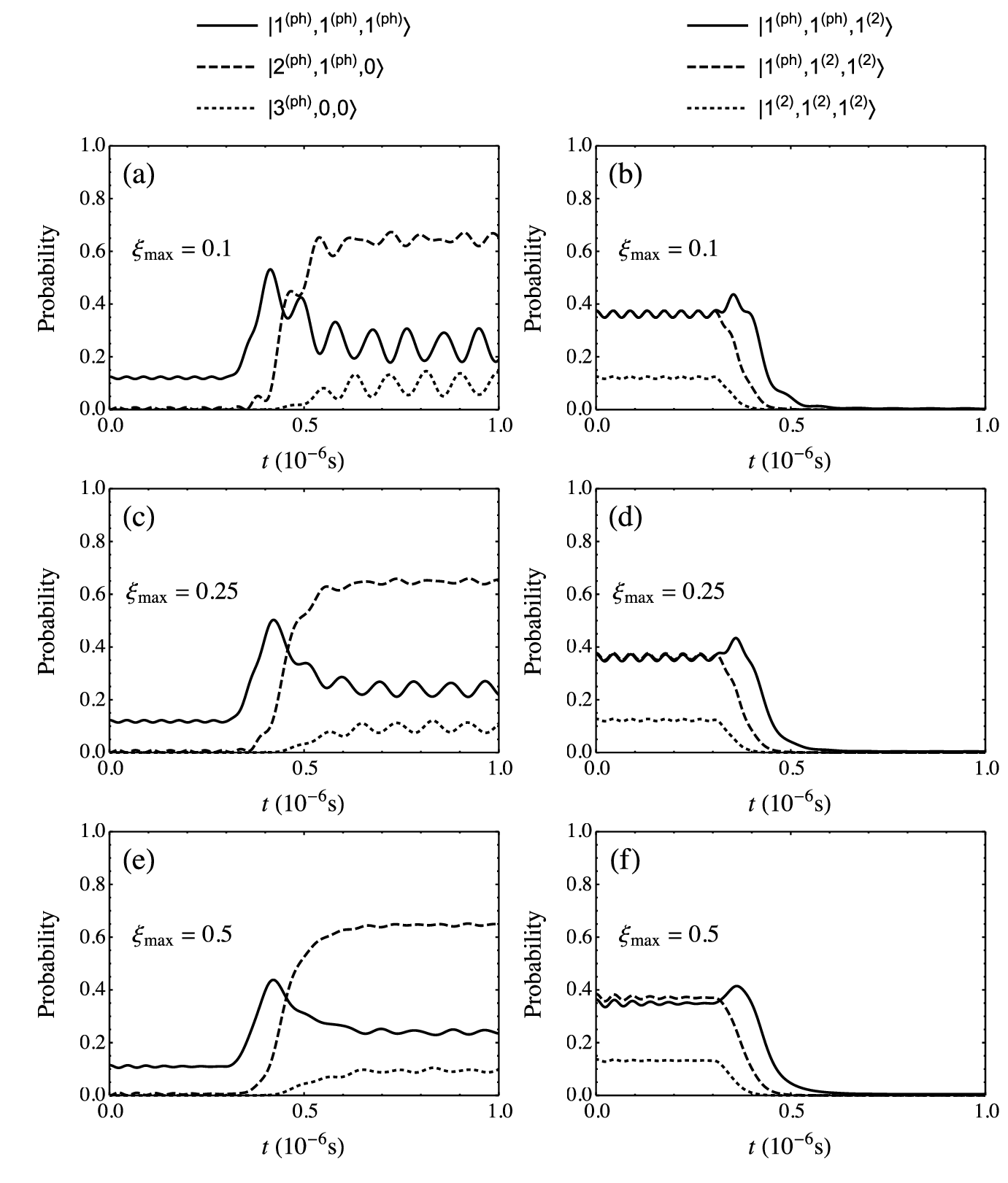}
	\end{center}
	\caption{The evolution of the average occupation probability of 6 classes of basis states in the wavefunction for uniform disorder with disorder strengths (a,b) 0.1, (c,d) 0.25 and (e,f) 0.5.\label{fig:popsplot}}
\end{figure*}

\Figref{popsplot} shows the results for the occupation probabilities of the basis states averaged over the 400 realizations of uniform disorder, with the states labelled in the same way as in the absence of disorder. Comparing the results without disorder in \Figref{popsplot0} to the results with uniform disorder in \Figref{popsplot} shows that the behaviour of the occupation probabilities of the states with disorder is similar to the behaviour in the absence of disorder. However, the amplitude of oscillations in the occupation probabilities decreases as the strength of the disorder increases, as was the case for the correlation functions. Additionally, the initial maximum in the average occupation probability of the state with 1 photon in each cavity decreases as the strength of the disorder increases.

\subsection{Speckle Disorder\label{sec:speckle}}

In addition to the simulations above using uniform disorder, we also ran simulations where speckle disorder was added to the Rabi frequency. For these simulations we used the same values for $g_{13}$, $g_{24}$, $\varepsilon$, $\delta$, $\Delta$, $N$ and $2\omega\alpha$ as in section \ref{sec:clean}. We used 400 realizations of speckle disorder, and this number was found to be sufficient for the same reasons as in the case of uniform disorder.

\begin{figure*}
	\begin{center}
		\includegraphics[width=0.95\textwidth]{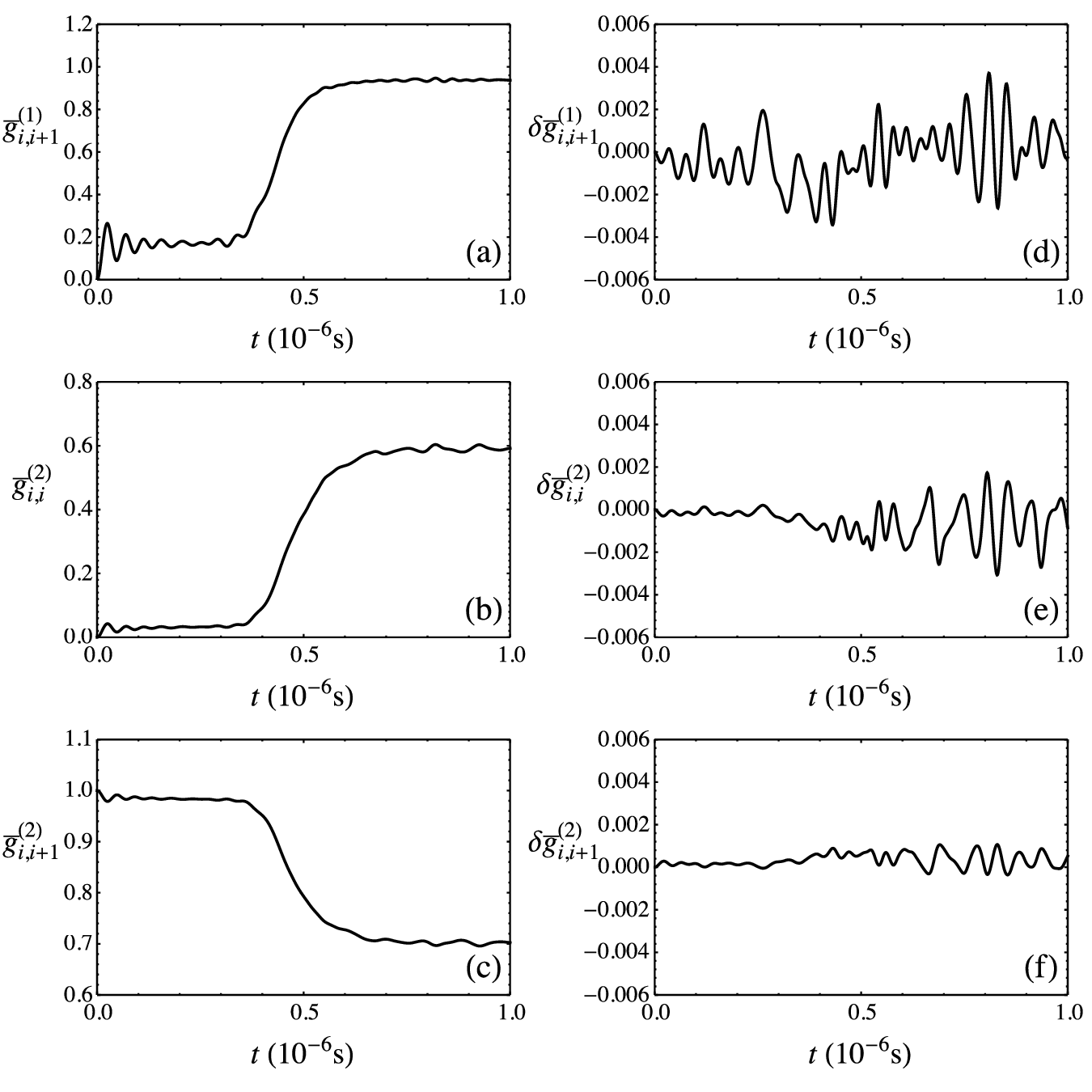}
	\end{center}
	\caption{The (a) normalized first-order correlation, (b) normalized on-site second-order correlation function and (c) normalized inter-cavity second-order correlation function averaged over all 3 cavities and 400 realizations of speckle disorder for the full Hamiltonian. Also, the difference in the (d) normalized first-order correlattion function, (e) normalized on-site second-order correlation function and (f) normalized inter-cavity second-order correlation function between the full and Bose-Hubbard Hamiltonians.\label{fig:spfulldiffplot}}
\end{figure*}

\Figsref{spfulldiffplot} (a), (b) and (c) show the results for the normalized first-order correlation function, the normalized on-site second-order correlation function and the normalized inter-cavity second-order correlation function when averaged over the 400 realizations of speckle disorder. \Figsref{spfulldiffplot} (d), (e) and (f) show the corresponding differences between the results for the full and Bose-Hubbard Hamiltonians. The magnitude of the disorder in the Rabi frequency for speckle disorder is large, and so the oscillations in the first- and second-order correlation functions are heavily damped. For the latter half of the time evolution, the maximum percentage differences between the results for the full Hamiltonian and the Bose-Hubbard Hamiltonian are about 0.004\%, 0.005\% and 0.002\% for the normalized first-order correlation function, the normalized on-site second-order correlation function and the normalized inter-cavity second-order correlation function respectively. These are much smaller than the values obtained in the cases of no disorder and uniform disorder.

\begin{figure*}
	\begin{center}
		\includegraphics[width=0.95\textwidth]{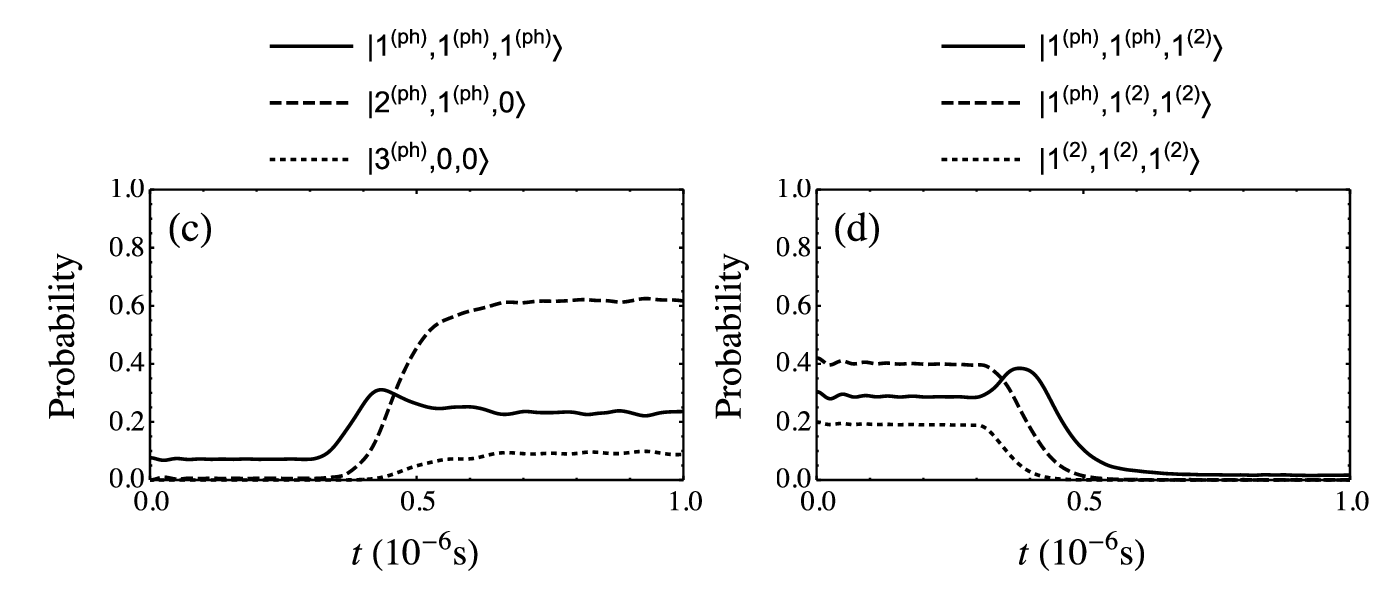}
	\end{center}
	\caption{The evolution of the occupation probability of 6 classes basis states in the wavefunction averaged over 400 realizations of speckle disorder.\label{fig:sppopsplot}}
\end{figure*}

The evolution of the average occupation probabilities of the 6 classes of basis states is shown in \Figsref{sppopsplot}. The oscillations in the probability values are even more damped than they are in the case of uniform disorder. Also, the initial maximum in the occupation probability of the state with 1 photon in each cavity is greatly reduced.

\subsection{Dipole Couplings\label{sec:dipole}}

\subsubsection{Correlation Functions\label{sec:dipolecorr}}

We next examined how varying the dipole couplings in the absence of disorder affects the correlation functions. We used the values $g_{13} = g_{24} = \quantity{1.0}{9}{\per\second}$, $\quantity{1.5}{9}{\per\second}$, $\quantity{2.0}{9}{\per\second}$, and $\quantity{2.5}{9}{\per\second}$. The parameters $\varepsilon$, $\delta$, $\Delta$, $N$ and $2\omega\alpha$ had the same values as in section \ref{sec:clean}. We used the same evolution of the Rabi frequency and started with the same initial state of 1 polariton in each cavity.

\begin{figure*}
	\begin{center}
		\includegraphics[width=0.95\textwidth]{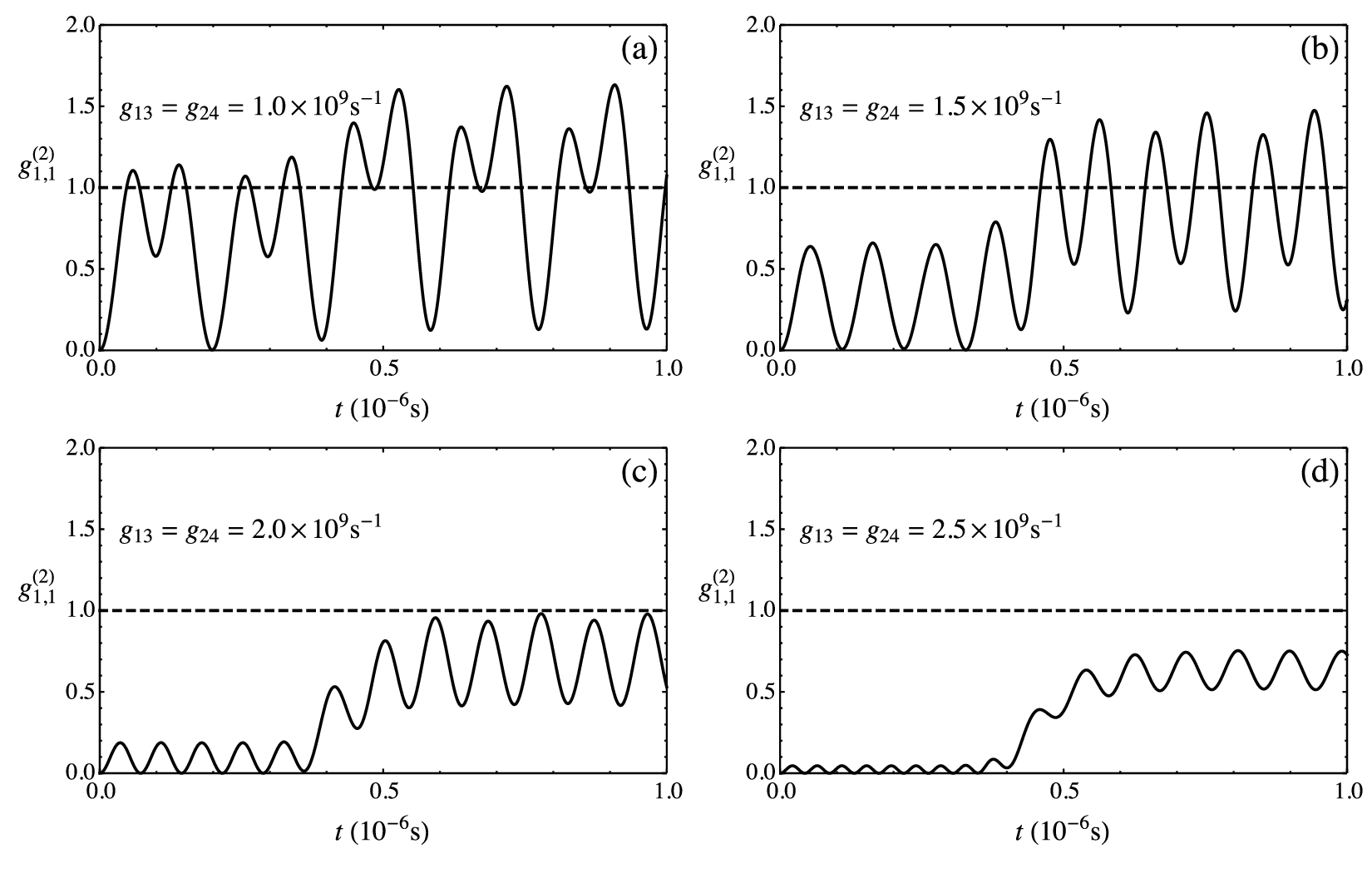}
	\end{center}
	\caption{The normalized on-site second-order correlation in cavity 1 for the four dipole couplings.\label{fig:gg211plot}}
\end{figure*}

\begin{figure*}
	\begin{center}
		\includegraphics[width=0.95\textwidth]{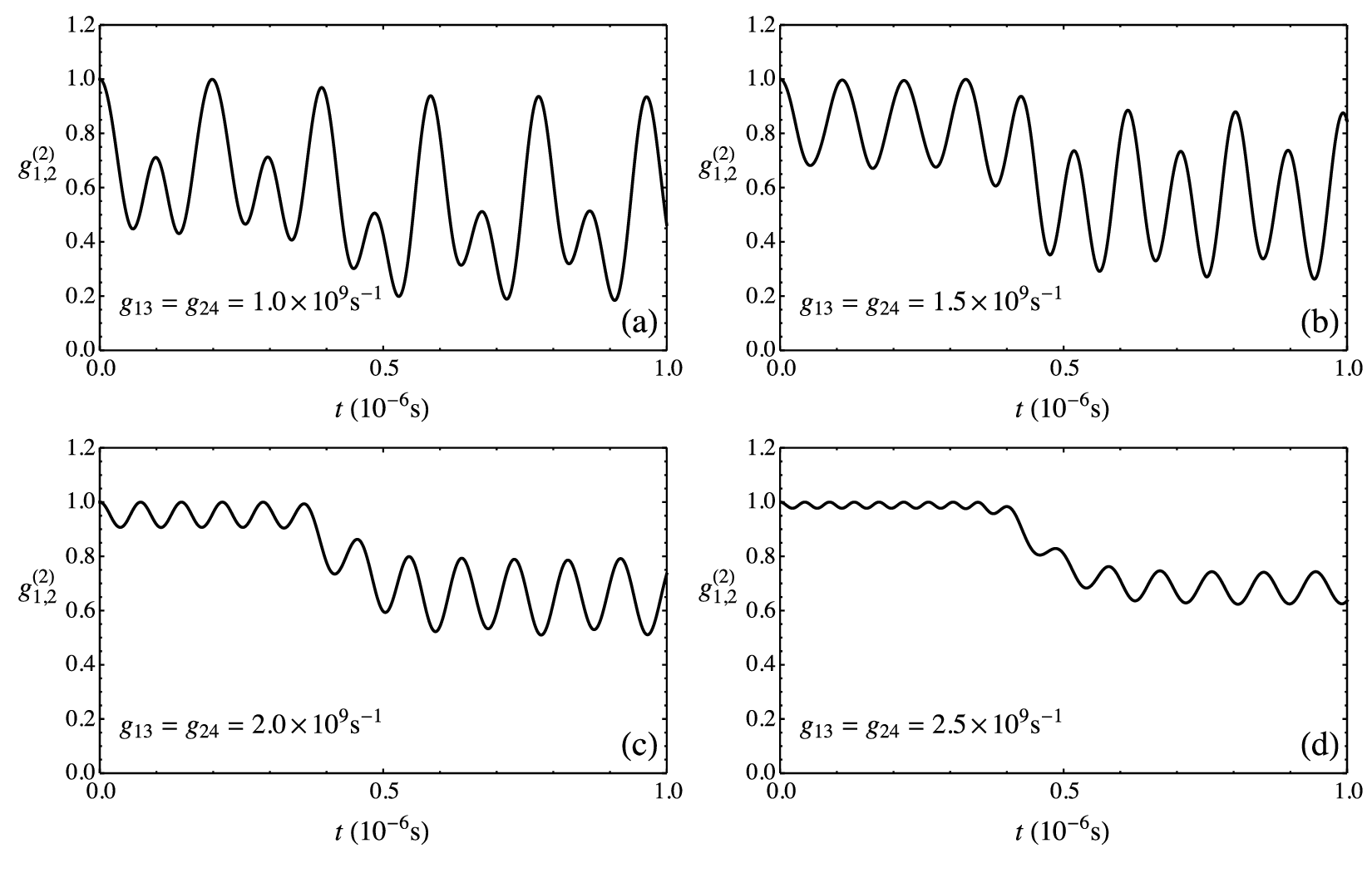}
	\end{center}
	\caption{The normalized inter-cavity second-order correlation between cavities 1 and 2 for the four dipole couplings.\label{fig:gg212plot}}
\end{figure*}

\Figref{gg211plot} shows the results for the normalized on-site second-order correlation function for the four dipole couplings. The magnitude of the oscillations in the correlation functions decreases as the dipole coupling increases. The initial number fluctuations (which are equal to the normalized on-site second-order correlation functions) are large in \Figref{gg211plot} (a) and (b), indicating a superfluid initial phase for the two smaller dipole couplings. Meanwhile the number fluctuations in \Figref{gg211plot} (c) and (d) are smaller, indicating a Mott-insulator initial phase for the two larger dipole couplings. Also, for the two smaller dipole couplings, the on-site second-order correlation functions oscillate between antibunching ($g_{11}^{(2)} < 1$) and bunching ($g_{11}^{(2)} > 1$) of the polaritons. \Figref{gg212plot} shows the results for the normalized inter-cavity second-order correlation function. Again the magnitude of the oscillations in the correlation functions decreases as the dipole coupling increases.

\begin{figure*}
	\begin{center}
		\includegraphics[width=0.95\textwidth]{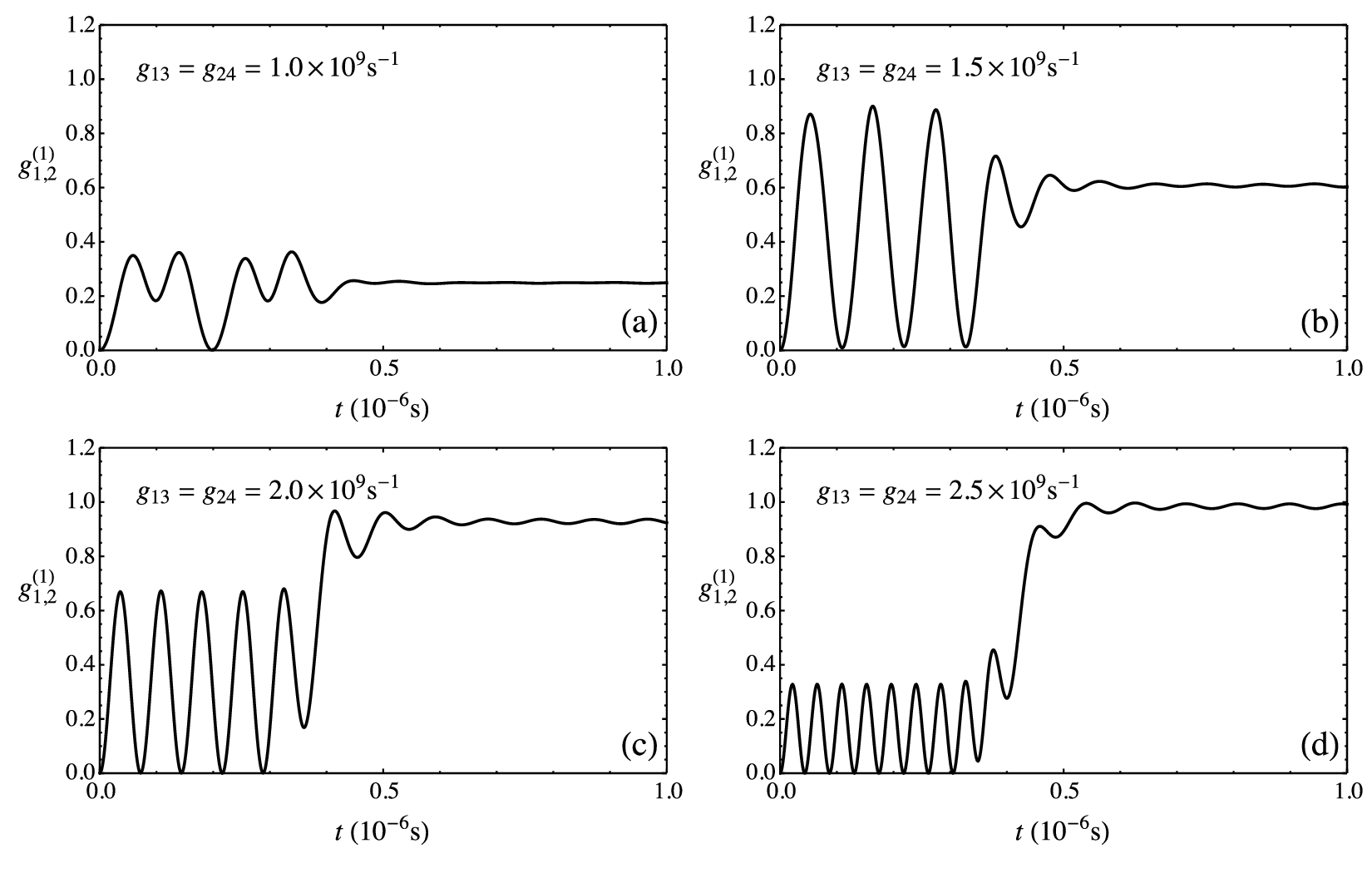}
	\end{center}
	\caption{The normalized first-order correlation between cavities 1 and 2 for the four dipole couplings.\label{fig:gg1plot}}
\end{figure*}

\Figref{gg1plot} shows the results for the normalized first-order correlation function for the four values of the dipole coupling. In each case, the correlations start off with larger oscillations and then the oscillations greatly decrease as the Rabi frequency is increased. The final values of the normalized first-order correlation function for the four couplings, $\quantity{1.0}{9}{\per\second}$, $\quantity{1.5}{9}{\per\second}$, $\quantity{2.0}{9}{\per\second}$, and $\quantity{2.5}{9}{\per\second}$, are $g_{12}^{(1)} \approx$ 0.25, 0.60, 0.92, and 0.99 respectively. The final value therefore increases as the dipole coupling increases and the initial phase goes from superfluid to Mott-insulator, with almost full coherence only occurring for the largest dipole coupling.

\subsubsection{Occupation Probabilities\label{dipoleprob}}

\begin{figure*}
	\begin{center}
		\includegraphics[width=0.95\textwidth]{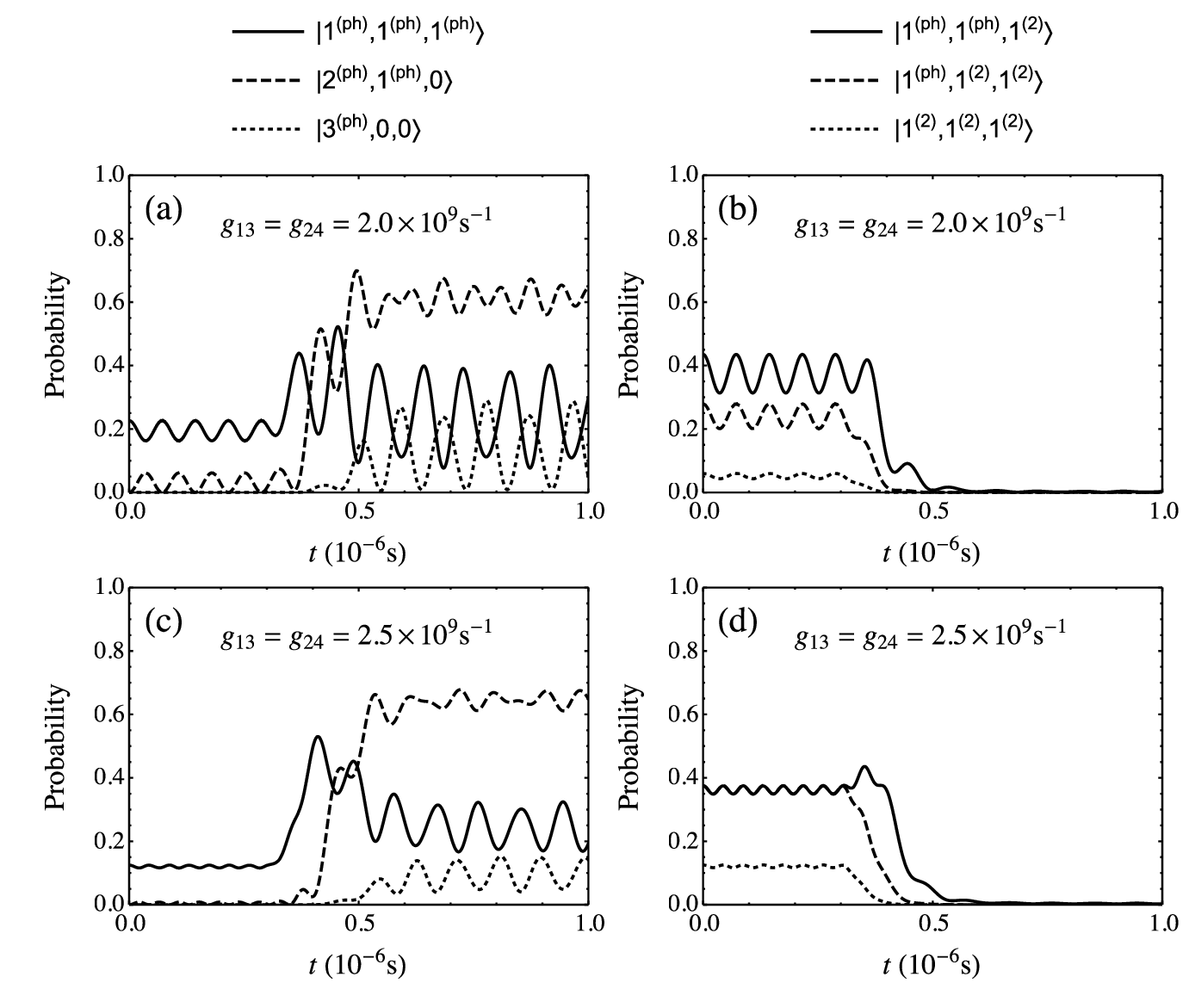}
	\end{center}
	\caption{The evolution of the average occupation probability of 6 classes of basis states for the dipole couplings (a,b) $\quantity{2.0}{9}{\per\second}$ and (c,d) $\quantity{2.5}{9}{\per\second}$.\label{fig:pops34plot}}
\end{figure*}

Next we examine the occupation probabilities of various basis states for the four dipole couplings. We use the same parameters as in the previous section and the same 6 classes of basis states as before. \Figref{pops34plot} shows the results for the dipole couplings $\quantity{2.0}{9}{\per\second}$ and $\quantity{2.5}{9}{\per\second}$. Compared to the results in \Figref{pops34plot} (a) and (b), the time evolution of the occupation probabilities in \Figref{pops34plot} (c) and (d) shows larger oscillations. After $0.5\,\mu\second$, the mixed states have negligible contribution and among the photonic states, the $\ket{2^{(\mathrm{ph})},1^{(\mathrm{ph})},0}$ states dominate.

\begin{figure*}
	\begin{center}
		\includegraphics[width=0.95\textwidth]{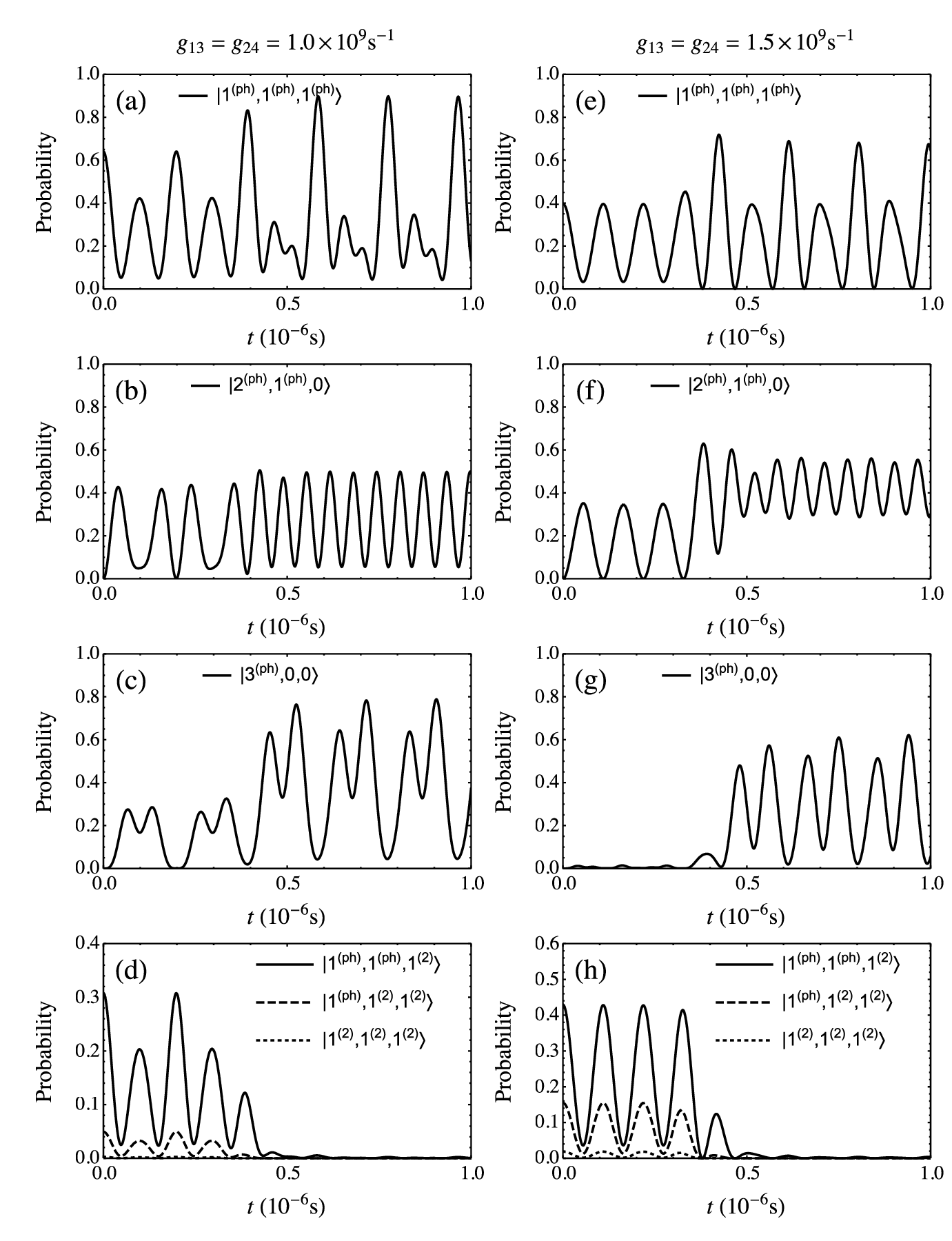}
	\end{center}
	\caption{The evolution of the average occupation probability of 6 classes of basis states for the dipole couplings (a-d) $\quantity{1.0}{9}{\per\second}$ and (e-h) $\quantity{1.5}{9}{\per\second}$.\label{fig:pops12plot}}
\end{figure*}

\Figref{pops12plot} shows the results for dipole couplings $\quantity{1.0}{9}{\per\second}$ and $\quantity{1.5}{9}{\per\second}$. For both of these two smaller dipole couplings, the occupation probabilities show larger oscillations than for the two larger couplings. In particular, the periodically large occupation probability of the $\ket{3^{(\mathrm{ph})},0,0}$ states for the two smaller dipole couplings coincides with the occasional bunching seen in the on-site second-order correlation functions for these dipole couplings.

\section{Conclusion\label{sec:conclusion}}

In this paper we investigated the dynamics of a coupled array of optical cavities, each containing an ensemble of atoms driven by an external laser, when disorder is introduced into the intensity of the driving laser. The simulations were performed in the absence of disorder and then for both uniform disorder and speckle disorder in the Rabi frequency of the driving by the external laser. The mean Rabi frequency was increased from a value of $\quantity{7.9}{10}{\per\second}$, which would put the system in the Mott-insulator phase, up to a value of $\quantity{1.1}{12}{\per\second}$, which would put it in the superfluid phase. We examined the evolution of the normalized first- and second-order correlation functions for the full Hamiltonian and the approximate Bose-Hubbard Hamiltonian that describe the system. We also investigated the evolution of the photon number, collective level 2 atomic excitation number and the occupation probabilities of certain basis states for the full Hamiltonian.

Both with and without disorder, the normalized first-order correlation function was found to start off near 0 and then increase as the Rabi frequency was increased to a value near 1. Also, the normalized on-site second-order correlation function was found to start off near 0 and then increase to a value less than 1, indicating antibunching of the polaritons. Meanwhile the normalized inter-cavity second-order correlation function was found to start off at a value of 1 and then decrease. All three quantities showed oscillations superimposed on the overall increase or decrease, and the magnitude of the oscillations decreased as the strength of the disorder increased. For all of the quantities, the results for the full and Bose-Hubbard Hamiltonians matched well and matched more closely as the strength of the disorder increased.

In the absence of disorder, the photon number and collective level 2 atomic excitation number were found to start off at 0.5. As the Rabi frequency was increased, the photon number increased to 1 and the atomic excitation number decreased to 0, indicating that the polaritons became entirely photonic.

Without disorder, the initial occupation probability of basis states with 1 photon/collective level 2 atomic excitation in each cavity was non-zero, with other states having zero probability. As the Rabi frequency was increased, the occupation probabilities of the mixed states decreased to 0 after approximately $0.5\,\mu\second$, while the occupation probabilities of purely photonic states increased. The occupation probability of the state with 1 photon in each cavity initially increased to a maximum and then decreased. All the occupation probabilities showed oscillations superimposed on the general increase or decrease. With the addition of disorder, the magnitude of the oscillations in the occupation probabilities was increasingly damped as the strength of the disorder increased. Also the initial maximum in the probability of the state with 1 photon in each cavity decreased in value.

For all the correlation functions and occupation probabilities studied, the addition of disorder did not change the overall behaviour of the evolution, only the magnitude of the oscillations that occur on top of the general trend.

We then investigated the effect of using different dipole couplings for the levels 1-3 and 2-4 transitions. The normalized first-order correlation functions for the dipole couplings $\quantity{1.0}{9}{\per\second}$, $\quantity{1.5}{9}{\per\second}$ and $\quantity{2.0}{9}{\per\second}$ increased to values of only 0.25, 0.60 and 0.92 respectively. Therefore the system did not reach full coherence at the end of the evolution. On the other hand, for the dipole coupling $\quantity{2.5}{9}{\per\second}$, the system reached almost full coherence with the correlation function having a final value of 0.99. Also, the oscillations in the normalized on-site second-order correlation functions increased as the dipole coupling decreased, with the oscillations being large enough that the correlation function periodically exceeds 1 for the dipole couplings $\quantity{1.0}{9}{\per\second}$ and $\quantity{1.5}{9}{\per\second}$. Therefore, there was some polariton bunching in these two cases.

The oscillations in the occupation probabilities of the basis states generally increased as the dipole coupling decreased. The oscillations were large enough for the dipole couplings $\quantity{1.0}{9}{\per\second}$ and $\quantity{1.5}{9}{\per\second}$ that the occupation probabilities of states with non-uniform photon numbers in the cavities periodically got large enough to cause the bunching found in the correlation functions.

\section*{Author contribution statement}

Abuenameh Aiyejina performed the simulations. Both authors wrote the manuscript together.

\bibliographystyle{epj}
\bibliography{paper}

\end{document}